\RequirePackage{fix-cm}
\documentclass[smallextended]{svjour3}       
\smartqed  
\usepackage{graphicx}
\usepackage{amsmath, amssymb}
\usepackage{color}
\newcommand{\be}{\begin{equation}}  
\newcommand{\ee}{\end{equation}}
\newcommand{\beq}{\begin{eqnarray}}  
\newcommand{\eeq}{\end{eqnarray}}

\def\ud{{\mathrm{d}}}
\def\udl{\stackrel{\leftarrow}{\ud}}

\def\im{{\mathrm{i}}}
\def\ex{{\mathrm{e}}}

\def\bf{\mbox{\boldmath $f$}}
\def\bg{\mbox{\boldmath $g$}}
\def\bA{\mbox{\boldmath $A$}}

\def\bG{\mbox{\boldmath $G$}}  
\def\bH{\mbox{\boldmath $h$}}  
\def\bK{\mbox{\boldmath $K$}}

\def\bS{\mbox{\boldmath $S$}}

\def\unit{\mbox{\boldmath $1$}}

\def\bgG{\mbox{\boldmath $\varGamma$}}  
  
\def\bgS{\mbox{\boldmath $\varSigma$}}  
\def\a{\alpha}  
\def\b{\beta}  
\def\g{\gamma}

\def\eps{\epsilon}

\def\heff{\bH_{\text{eff}}}
\def\heffd{\bH_{\text{eff}}^{\dagger}}

\begin{document}

\title{Formal equivalence between partioned and partition-free quenches in quantum transport}

\author{Michael Ridley         \and
        Riku Tuovinen 
}


\institute{Michael Ridley \at
              The Raymond and Beverley Sackler Center for Computational Molecular and Materials Science, Tel Aviv University, Tel Aviv 6997801, Israel \\
              \email{ridley@mail.tau.ac.il}           
           \and
           Riku Tuovinen \at
              Max Planck Institute for the Structure and Dynamics of Matter, Center for Free Electron Laser Science, 22761 Hamburg, Germany \\
              \email{riku.tuovinen@mpsd.mpg.de}
}

\date{Received: date / Accepted: date}

\maketitle

\begin{abstract}
In this paper we review the partitioned and partition-free approaches to the calculation of the time-dependent response of a molecular junction to the switch-on of an arbitrary time-dependent bias. Using the non equilibrium Green's function formalism on different time contours, we derive a formal equivalence between these two approaches. This clarifies a recent result of [PRB 95, 104301 (2017)], which is valid for a static bias and single level molecular structure, and extends it to arbitrary time-dependent biases and arbitrarily large molecular structures.
\end{abstract}

\section{Introduction}
\label{intro}
With the advent of molecular devices that can operate in the GHz-THz regime, the field of molecular electronics is reaching the point of technological relevance. Examples of time-dependent nanoelectrical components include frequency doublers and detectors  ~\cite{Iniguez-de-la-Torre2010}, stereoelectric ~\cite{su2015} and photo-induced ~\cite{jia2016}  switches, and AC-driven graphene nanoribbon field-effect transistors ~\cite{zhang2014}. Concomitantly with these experimental developments, a flurry of sophisticated theoretical approaches to the computation of transport quantities in molecular junctions has emerged. All such methods involve the computation of time-dependent ensemble averages in response to a sudden change in the Hamiltonian describing the molecular junction. This sudden change can be introduced in several different ways, depending on which kind of quench or switch-on process one wishes to simulate. The quench or switch-on time $t_0$ marks the beginning of an irreversible transport process.

\begin{figure}[t]
\centering
\includegraphics[width=0.75\textwidth]{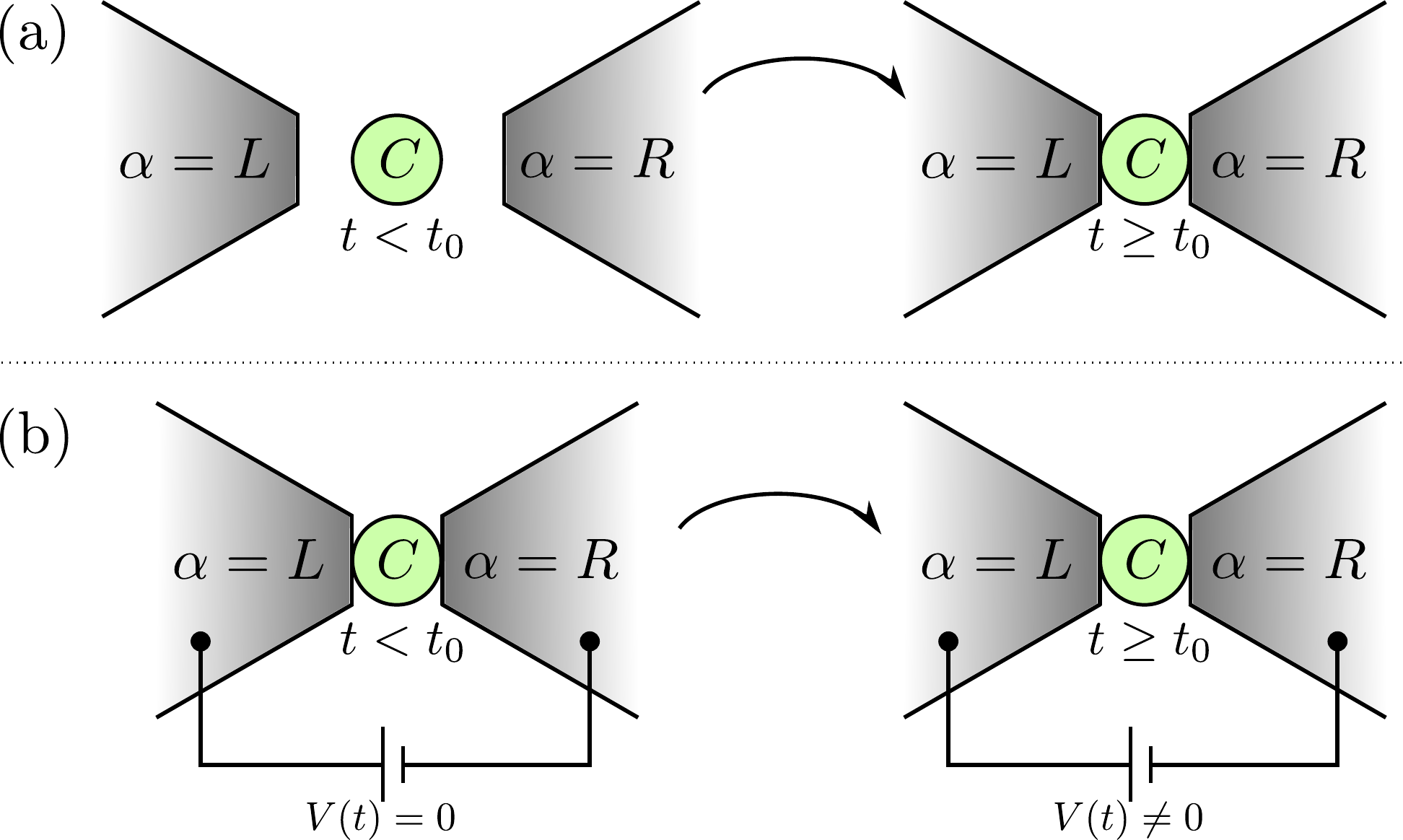}
\caption{Schematic representations of the transport setup with switch-on time $t_0$. (a) Partitioned approach: For $t<t_0$ the leads are disconnected from the central region, and they are in separate thermodynamical equilibria (different temperatures and chemical potentials). Once the contact is established for $t\geq t_0$ charge carriers start to flow. (b) Partition-free approach: The whole system is for $t<t_0$ in a global thermodynamical equilibrium (unique temperature and chemical potential). For $t\geq t_0$ a bias is applied in the leads and charge carriers start to flow.}
\label{fig:partition}
\end{figure}

Historically, most first-principles calculations of the current in a nano sized conductor derived from the method of Landauer and B{\"u}ttiker~\cite{Landauer1970,Buttiker1986}. This approach provides an 
intuitive physical picture of electron transport: the current $I_{\a\b}$ in lead $\b$, is calculated from the 
scattering states originating from lead $\a\neq\b$. This is typically written in terms of transmission 
probabilities for an electron to traverse from lead $\a$ to lead $\b$. Summing over all leads 
$\a\neq\b$ the difference $I_{\a\b}-I_{\b\a}$ of the currents flowing into and out of lead $\b$ then 
gives the \emph{stationary} current $I_\b$ in terminal $\b$. The calculation of \emph{transient} 
current is more intricate. In the transient regime it is important to make the distinction
between \emph{partitioned} and \emph{partition-free} approaches, meaning whether the conducting
device is disconnected from the lead environment in equilibrium (partitioned) or whether the whole
system is initially contacted in a unique equilibrium (partition-free). This is depicted in 
Fig.~\ref{fig:partition}. A microscopic derivation, based on the time-dependent Schr{\"o}dinger
equation, for the transient current was presented by Caroli et al.~\cite{caroli1,caroli2} who 
considered a partitioned approach. The Landauer--B{\"u}ttiker formula is recovered in the limit 
$t\to\infty$. Further work using the partitioned approach was put forward by Meir, Wingreen and Jauho~\cite{MeirWingreen1992,Jauho1994}. An alternative approach was provided by Cini~\cite{cini} who considered the partition-free approach. Although the initial state was different in Cini's approach, the same Landauer--B{\"u}ttiker formula for
the current is recovered at the long-time limit. This indicated that the system preparation has no effect on the steady-state properties, a fact which was later clarified and rigorously proven by Stefanucci and Almbladh in their so-called Theorem of Equivalence~\cite{Stefanucci2004}. More detailed connections between partitioned and partition-free approaches have been considered in Refs.~\cite{Stefanucci2004EPL,Stefanucci2006}. In these works it is also pointed out that the partition-free approach can be directly connected with time-dependent density-functional theory as the voltage switch-on is a local potential linearly coupled to the particle density; this is not possible with the partitioned approach where the coupling process is not local in space. In the present paper we will prove a new relationship between the ensemble averages computed from the different switch-on methods, which goes beyond the Theorem of Equivalence, in that it is also valid in the transient regime. 

The Nonequilibrium Green's function (NEGF) formalism~\cite{svlbook} provides a natural framework to calculate the current at \emph{all times}, and it is not limited to the steady state. A calculation of $I_\b(t)$ was done by 
Jauho et al.~\cite{Jauho1994} where the partitioned approach was used to write $I_\b(t)$ as a double 
integral over time and energy. When the leads are described within the wide-band limit approximation (WBLA), it is 
possible to perform the time integral analytically and obtain a time-dependent extension to the 
Landauer--B{\"u}ttiker formula, which we refer to in this paper as the TD-LB formalism~\cite{Stefanucci2004,Perfetto2008,Tuovinen2013,Tuovinen2014,Ridley2015,ridleytuovinen2017}.
In these works, the approach of Cini's is used, without partitioning the transport setup, and thus 
the loss of memory from the initial state is confirmed. In addition, the practical usefulness of 
a ``Landauer--B{\"u}ttiker formula'' is retained: for a wide class of time-dependent driving fields ~\cite{ridleytuovinen2017,Ridley2016a,Ridley2016b}, the calculation of the transient current for each
value of time $t>t_{0}$ is no more computationally expensive than the calculation of the steady-state. 
Recently, a work was published that examines the relationship between the partitioned and partition-free approaches in the wide-band limit approximation (WBLA) ~\cite{Odashima2017}. They showed that, when the bias switched on at the beginning of a transport experiment is static, and the molecule sandwiched between the conducting leads has only a single level, then it is possible to exactly map from the partitioned approach of Jauho \emph{et al.} to the partition-free TD-LB formalism developed by the present authors. In this work, we extend the validity of this result to include arbitrarily complex molecular structures and arbitrarily time-dependent biases. 

\section{The NEGF formalism}
\begin{figure}[t]
\centering
\includegraphics[width=0.75\textwidth]{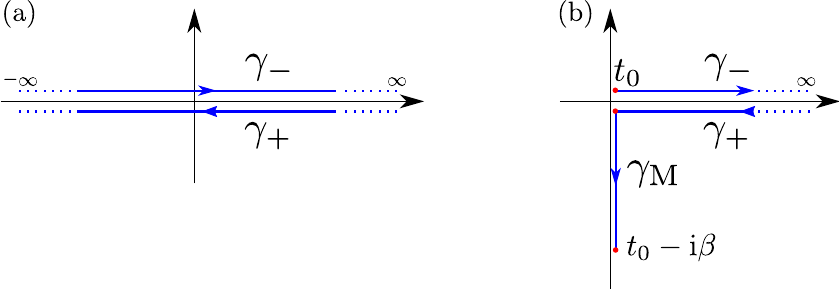}
\caption{(a) Two-branch time-ordered Keldysh contour (b) Konstantinov--Perel' contour has a forward ($\gamma_-$) and a backward branch ($\gamma_+$) on the real-time axis, $[t_0,\infty[$ and a vertical branch on the imaginary axis $\gamma_{\text{M}}$, $[t_0,t_0-\im\beta]$ with inverse temperature $\beta$. In both contour-orderings the arguments on the upper branch appear earlier even though the real-time projections could be larger. (The contours are slightly shifted from the axes for clarity.)}
\label{fig:timecontour}
\end{figure}

The Hamiltonian for our transport setup in Fig.~\ref{fig:partition} is (for times $z$ on the chosen contour of Fig.~\ref{fig:timecontour})
\beq
\hat{H}\left(z\right) & = & \underset{k\alpha\sigma}{\sum}\varepsilon_{k\alpha}\left(z\right)\hat{d}_{k\alpha\sigma}^{\dagger}\hat{d}_{k\alpha\sigma}+\underset{mn\sigma}{\sum}h_{mn}\left(z\right)\hat{d}_{m\sigma}^{\dagger}\hat{d}_{n\sigma} \nonumber \\
& + &\underset{m k\alpha\sigma}{\sum}\left[T_{mk\alpha}\left(z\right)\hat{d}_{m\sigma}^{\dagger}\hat{d}_{k\alpha\sigma}+T_{k\alpha m}\left(z\right)\hat{d}_{k\alpha\sigma}^{\dagger}\hat{d}_{m\sigma}\right] . \label{eq:Hamiltonian}
\eeq
The first term accounts for the $\a$-th lead with $k\a$ being the $k$-th basis function of the $\a$-th lead, the second term is for the central conducting device, and the last term is for the coupling between the central part and the $\a$-th lead. The corresponding annihilation (creation) operators for these states are denoted by $\hat{d}^{(\dagger)}$ obeying the fermionic anticommutation rules $\{\hat{d}_{x\sigma},\hat{d}_{y\sigma'}^\dagger\} = \delta_{xy}\delta_{\sigma\sigma'}$ for indices $x,y$ belonging either to the leads or to the central region. We exclude interactions (electron--electron, electron--phonon) in our model for we wish to consider analytically solvable systems. In principle, this formulation could also incorporate interactions but in practice this can be computationally very demanding.

Depending on the partitioning scheme in Fig.~\ref{fig:partition} we prepare our transport setup in either of the depicted descriptions:
\begin{itemize}
\item[(a)] In the remote past ($t\to-\infty$) leads are disconnected from the central region and in separate equilibria at different chemical potentials $\mu_L$, $\mu_R$ and different inverse temperatures $\beta_L$,$\beta_R$. The density matrix for the partitioned system could then be written as $\hat{\rho}=\ex^{-\beta_L\sum_k (\eps_{kL} - \mu_L)\hat{d}_{kL}^\dagger\hat{d}_{kL}} \times \hat{\rho}_C \times \ex^{-\beta_R\sum_k (\eps_{kR} - \mu_R)\hat{d}_{kR}^\dagger\hat{d}_{kR}}$. At a particular time, which we refer to as the \emph{coupling time} $t_{c}$, the leads are connected to the molecule. If there is a temperature and/or chemical potential gradient across the system as a result of this coupling, charge will start to flow. In addition, there is a bias switch-on time $t_{0}$ which may or may not coincide with $t_{c}$. At times $t>t_0$ the system is driven out of equilibrium with an external field by changing the energy levels of every lead $\varepsilon_{k\alpha}\to \varepsilon_{k\a}+V_\a(t)$.
\item[(b)] The whole system is initially ($t<t_0$) in a global thermal equilibrium at inverse temperature $\b$ and in chemical potential $\mu$. The density matrix is of the form $\hat{\rho} = \frac{1}{\mathcal{Z}}\ex^{-\b(\hat{H} -\mu\hat{N})}$ where $\hat{H}$ is the total Hamiltonian and $\hat{N}$ and $\mathcal{Z}$ are the particle number operator and the grand-canonical partition function, respectively. At times $t>t_0$ the energy levels of every lead are shifted with a spatially homogeneous shift $\varepsilon_{k\alpha}\to \varepsilon_{k\a}+V_\a(t)$.
\end{itemize}
For both cases (a) and (b) charge carriers will flow through the central molecule after $t=t_0$ due to the nonequilibrium conditions. In case (b), one could have a situation in which the coupling and bias switch-on times are different, i.e. $t_{c} \neq t_{0}$, although in the formulation of Ref. ~\cite{Jauho1994} they are chosen to be identical. In the case of $t_{c} \neq t_{0}$, a net current flows after $t_{c}$ unless $\beta_L=\beta_R$ and $\mu_L=\mu_R$. 
 The two kinds of quench are typically represented using the two different contours shown in Fig.~\ref{fig:timecontour}. Fig.~\ref{fig:timecontour} (a) shows the Keldysh contour ~\cite{Keldysh1965}, composed of upper ($\gamma_-$) and lower ($\gamma_+$) branches of real times. Note that the upper limits of this contour may be chosen at either the observation time $t$ or $+\infty$ ~\cite{svlbook}. The Keldysh contour is usually used for the partitioned approach, as the way in which the system was prepared prior to the coupling quench is relegated to the distant past ~\cite{Jauho1994}. When the Keldysh contour is used, we can define the Hamiltonian in Eq. (\ref{eq:Hamiltonian}) for all contour times as follows:
\beq
\varepsilon_{k\a}(z) & = & \begin{cases}\varepsilon_{k\a}  + V_\a(t)  & \text{when} \ z\equiv t \geq t_0 \\ \varepsilon_{k\a}& \text{when} \ z\equiv t<t_0\end{cases} , \\
h_{mn}(z) & =& \begin{cases} h_{mn}+u_{mn}+V_{C}(t)\delta_{mn} & \text{when} \  z\equiv t \geq t_0\\ h_{mn} & \text{when} \ z\equiv t<t_0 \end{cases} \label{eq:HzCC1}, \\
T_{k\a m}(z) & = & \begin{cases} T_{k\a m} & \text{when} \ z\equiv t\geq t_c \\ 0 & \text{when} \ z\equiv t <t_c \end{cases}
\eeq

In Fig.~\ref{fig:timecontour} (b) the Konstantinov--Perel' (KP) contour ~\cite{KonstantinovPerel1961} is displayed, which consists of an upper brunning from $t_{0}$ to the observation time $t$ and vice versa, in addition to a vertical Matsubara branch $\gamma_{\text{M}}$ which runs from $t_{0}$ to $t_0-\im\beta$. The Matsubara branch describes the system in equilibrium, and so the chemical potential is included in the Matsubara part of the Hamiltonian defined on $\gamma_{\text{M}}$:

\beq
\varepsilon_{k\a}(z) & = & \begin{cases}\varepsilon_{k\a} + V_{\alpha}(t) & \text{when} \ z \in \gamma_- \oplus \gamma_+ \\ \varepsilon_{k\a}-\mu & \text{when} \ z \in \gamma_{\text{M}}\end{cases} , \\
h_{mn}(z) & = & \begin{cases} h_{mn}+u_{mn}+V_{C}(t)\delta_{mn} & \text{when} \  z\in \gamma_-\oplus \gamma_+\\ h_{mn}-\mu\delta_{mn} & \text{when} \ z\in \gamma_{\text{M}} \end{cases} \label{eq:HzCC2}, \\
T_{k\a m}(z) & = & \begin{cases} T_{k\a m} & \text{when} \ z \in \gamma_- \oplus \gamma_+ \\ T_{k\a m} & \text{when} \ z \in \gamma_{\text{M}}\end{cases}
\eeq

To calculate time-dependent nonequilibrium quantities we use the equations of motion for the one-particle Green's function on the KP contour $\g$ of Fig.~\ref{fig:timecontour} (b). This is defined as the ensemble average of the contour-ordered product of particle creation and annihilation operators in the Heisenberg picture~\cite{svlbook}
\be\label{eq:greenf}
G_{rs}(z,z') =
-\im\langle\mathcal{T}_\g[\hat{d}_{r,\mathrm{H}}(z)\hat{d}_{s,\mathrm{H}}^\dagger(z')]\rangle,
\ee
where the indices $r$, $s$ can be either indices in the leads or in the central region and the variables $z$, $z'$ run on the chosen time contour. For a function $k(z,z')$ defined on the time contour we extract the following components: lesser ($<$), greater ($>$), retarded (r), advanced (a), left ($\lceil$), right ($\rceil$) and Matsubara (M)~\cite{svlbook}
\begin{subequations}
\beq
k^{\lessgtr}(t,t')&=&k(t_{\mp},t'_{\pm}), \label{eq:func-lss-gtr}\\
k^{{\rm r}/{\rm a}}(t,t')&=&\pm\theta(\pm(t-t'))\left[k^{>}(t,t')-k^{<}(t,t')\right], \label{eq:func-ret-adv}\\
k^{\lceil}(\tau,t')&=&k(t_{0}-\im\tau,t'),\label{eq:func-left}\\
k^{\rceil}(t,\tau)&=&k(t,t_{0}-\im\tau),\label{eq:func-right}\\
k^{\rm M}(\tau,\tau')&=&k(t_{0}-\im\tau,t_{0}-\im\tau').\label{eq:func-mats}
\eeq
\end{subequations}
We note that the components $k^{\lceil,\rceil,\text{M}}$ are only related to the KP contour in Fig.~\ref{fig:timecontour}(b).

Although the following is valid for arbitrary number of leads $\a$, we now look at a specific setup of dividing the system in a transport sense into left ($L$) and right ($R$) leads and to a central conducting device ($C$), see Fig.~\ref{fig:partition}. We assume that the basis of the studied $LCR$ system can be divided, and we let the single-particle Hamiltonian and the Green's function be of the block form
\be
\underline{\bH} = \begin{pmatrix}  {\bH}_{LL} & {\bH}_{LC} & 0 \\ {\bH}_{CL} &  {\bH}_{CC} & {\bH}_{CR} \\ 0 & {\bH}_{RC} & {\bH}_{RR}\end{pmatrix} , \qquad \underline{\bG} = \begin{pmatrix}  {\bG}_{LL} & {\bG}_{LC} & {\bG}_{LR} \\ {\bG}_{CL} &  {\bG}_{CC} & {\bG}_{CR} \\ {\bG}_{RL} & {\bG}_{RC} & {\bG}_{RR}\end{pmatrix} ,
\ee
where $(\bH_{\a \a'})_{kk'} = \delta_{\a \a'}\delta_{k k'}\varepsilon_{k\a}(z)$ corresponds to the leads, $(\bH_{\a C})_{km} = T_{k\a m}(z)$ is the coupling part, and $(\bH_{CC})_{mn} = h_{mn}(z)$ accounts for the central (molecular) region. The block structure of the Hamiltonian means that the leads are coupled only through the central region as the direct couplings between them are zero. However, the matrix $\underline{\bG}$ can still have nonvanishing entries everywhere.

We now wish to recast the dynamics of the entire nanojunction in terms of Green's functions for the molecular sub-region $C$. For a Green's function matrix $\bG_{CC}$, whose dimension is  given by the basis size of the central region, we write the equation of motion as~\cite{Tuovinen2013} (and we omit the subscript $CC$ from now on)
\beq
\left[\im \frac{\ud}{\ud z} - {\bH}(z)\right]{\bG}(z,z') & = & \delta(z,z'){\unit} + \int_\g \ud \bar{z} {\bgS}(z,\bar{z}){\bG}(\bar{z},z')
\label{eq:eom-left} , \\
{\bG}(z,z')\left[-\im \frac{\udl}{\ud z'} - {\bH}(z')\right] & = &
\delta(z,z'){\unit} + \int_\g \ud \bar{z} {\bG}(z,\bar{z}){\bgS}(\bar{z},z')
\label{eq:eom-right} ,
\eeq
with a boundary condition such that the Green's function is antiperiodic along the contour $\g$ (Kubo--Martin--Schwinger boundary conditions). On the right-hand side of Eqs.~\eqref{eq:eom-left} and~\eqref{eq:eom-right} we introduced the \emph{embedding self-energy kernel}
\be
\bgS(z,z') = \sum_\a \bgS_\a(z,z') = \sum_\a \bH_{C\a}(z)\bG_{\a\a}(z,z')\bH_{\a C}(z'),
\ee
where $\bG_{\a\a}$ is the lead Green's function, satisfying $\left[\im\frac{\ud}{\ud z} - \bH_{\a\a}(z)\right]\bG_{\a\a}(z,z') = \delta(z,z')\unit_{\a\a}$. The self-energy here is solely due to the coupling between the central region and the leads since our model excludes many-body interactions. The commonly used \emph{wide-band limit approximation} (WBLA) makes the (retarded/advanced) self-energy proportional to a delta function 
\be\label{eq:wbla}
\bgS_{\alpha}^{\text{r}/\text{a}}\left(t,\bar{t}\right)=\mp\frac{\im}{2}\bgG_{\alpha}\delta\left(t-\bar{t}\right)
\ee
allowing for a closed solution to the equations of motion~\cite{svlbook,Tuovinen2013}.

In Refs.~\cite{Ridley2016a,Ridley2016b} the equations of motion~\eqref{eq:eom-left} and~\eqref{eq:eom-right} were solved using the WBLA for the self-energy. This was done by using the Langreth rules for mapping between contour and real-time convolution integrals ~\cite{Langreth1976}, leading to a separate equation of motion for each of the greater, lesser, retarded, advanced, left, right and Matsubara components of the Green's function defined above. Although the Hamiltonians defined in Eqs. (\ref{eq:HzCC1}) and (\ref{eq:HzCC2}) can be solved exactly ~\cite{Ridley2016b}, we assume for simplicity that $V_{C}(t)=0$ and $u_{mn}=0$, in which case the greater and lesser Green's functions are known to be 
\beq
\bG^{\gtrless}\left(t_{1},t_{2}\right) & = & \mp \im\int\frac{\ud\omega}{2\pi}f\left(\mp\left(\omega-\mu\right)\right)\underset{\alpha}{\sum}\bS_{\alpha}\left(t_{1},t_{0};\omega\right)\bgG_{\alpha}\bS_{\alpha}^{\dagger}\left(t_{2},t_{0};\omega\right) \label{eq:greater/lesserGF}
\eeq
where we have defined the matrices $\bS_{\alpha}$ as follows
\begin{equation}
\bS_{\alpha}\left(t,t_{0};\omega\right)\equiv \ex^{-\im\heff\left(t-t_{0}\right)}\left[\bG^{\text{r}}\left(\omega\right)-\im\bK_{\alpha}\left(t,t_{0};\omega\right)\right] . \label{eq:GFgreatless}
\end{equation}
In this expression, we have defined the effective non-hermitian hamiltonian for the molecular region, $\heff = \bH-\im\sum_\a\bgG_{\alpha}/2$. Thus the eigenenergies of the molecular structure are renormalised by the fact that it is open to an `environment' consisting of the leads. We have also introduced the frequency-dependent retarded Green's function
\begin{equation}
\bG^{\text{r}}\left(\omega\right)=\left(\omega\unit - \heff \right)^{-1}
\end{equation}
whose hermitian conjugate is the advanced component $\bG^{\text{a}}\left(\omega\right)$~\cite{svlbook}. In addition, we introduce the matrix object
 \begin{equation}
\bK_{\alpha}\left(t,t_{0};\omega\right)=\int_{t_{0}}^{t}\ud\bar{t}\ex^{-\im\left(\omega\unit-\heff\right)\left(\bar{t}-t_{0}\right)}\ex^{-\im\psi_{\alpha}\left(\bar{t},t_{0}\right)} . \label{eq:K-integral}
\end{equation}
Here, the time-dependent bias $V_{\alpha}\left(t\right)$
of lead $\alpha$ enters into Eq. (\ref{eq:GFgreatless}) only via the phase factors 
\begin{equation}
\psi_{\alpha}\left(t_{1},t_{2}\right)\equiv\int_{t_2}^{t_1} \ud\tau\, V_{\alpha}\left(\tau\right)\label{eq:phasealpha}
\end{equation}
in the integrand of the $\bK_{\alpha}$ matrix.

\section{Equivalence between partitioned and partition-free approaches}\label{app:equiv}
In Ref.~\cite{Odashima2017}, it was shown that the current and lesser Green's
function formulas obtained in the static bias partition-free approach of Ref.
\cite{Tuovinen2013,Tuovinen2014} can be reproduced within the partitioned approach of \cite{Jauho1994},
for the case when the molecular region of the junction consists of only one energy level. We now generalize this result in two ways, by allowing the molecule
to be composed of arbitrarily many levels, and by allowing the bias
driving the system to have any time-dependence. In the original derivation of Eq. (\ref{eq:GFgreatless}), Kadanoff--Baym equations for the Green's function
were solved on a contour on which a vertical branch was used to model
the equilibration of the system prior to the switch-on time $t_{0}$ \cite{Ridley2015}.
This lead to an expression for the lesser Green's function containing
four terms,
\begin{equation}
\bG^{<}\left(t_{1},t_{2}\right)=\sum_{i=1}^{4} \bG_{\left(i\right)}^{<}\left(t_{1},t_{2}\right) .
\end{equation}
The first term is simply due to the initial condition:
\beq\label{eq:term1}
\bG_{\left(1\right)}^{<}\left(t_{1},t_{2}\right) & = & \ex^{-\im\heff\left(t_{1}-t_{0}\right)}\bG^{<}\left(t_{0},t_{0}\right)\ex^{\im\heffd\left(t_{2}-t_{0}\right)} \nonumber \\
& = & \ex^{-\im\heff\left(t_{1}-t_{0}\right)}\int\frac{\ud\omega}{2\pi}f\left(\omega-\mu\right)\sum_\alpha \im\bA_{\alpha}\left(\omega\right)\ex^{\im\heffd\left(t_{2}-t_{0}\right)} ,
\eeq
where $\bA_{\alpha}\left(\omega\right)=\bG^{\text{r}}\left(\omega\right)\bgG_{\alpha}\bG^{\text{a}}\left(\omega\right)$
is the spectral function. Following a line integral in the two-time plane \cite{Ridley2015}, the remaining terms in the lesser Green's function are given in terms of real and imaginary convolution integrals defined with the following notation~\cite{svlbook}:
\beq
{[}\bf\cdot\bg{]}\left(z_{1},z_{2}\right) & \equiv &\int_{t_{0}}^{\infty}\ud\bar{t}\bf\left(z_{1},\bar{t}\right)\bg\left(\bar{t},z_{2}\right) , \label{eq:realtimeconv} \\
{[}\bf\star\bg{]}\left(z_{1},z_{2}\right) & \equiv & -\im\int_{0}^{\beta}\ud\tau\bf\left(z_{1},\tau\right)\bg\left(\tau,z_{2}\right)\label{eq:imagtimeconv} , 
\eeq
for functions $\bf$ and $\bg$ defined on the KP contour. Two terms arise from convolutions taken along the Matsubara branch of the KP contour:
\beq
& & \bG_{\left(2\right)}^{<}\left(t_{1},t_{2}\right) \nonumber \\
& = & \im\ex^{-\im\heff\left(t_{1}-t_{0}\right)}\int_{t_0}^{t_2}\ud\bar{t}\ex^{\im\heff\left(t_{1}-t_{0}\right)}\left[\bG^{\rceil}\star\bgS^{\lceil}\right]\left(t_{1},\bar{t}\right)\ex^{-\im\heffd\left(\bar{t}-t_{0}\right)}\ex^{\im\heffd\left(t_{2}-t_{0}\right)} \nonumber \\
& = & -\ex^{-\im\heff\left(t_{1}-t_{0}\right)}\int\frac{\ud\omega}{2\pi}f\left(\omega-\mu\right)\sum_\alpha \bG^{\text{r}}\left(\omega\right)\bgG_{\alpha}\bK_{\alpha}^{\dagger}\left(t_{2},t_{0};\omega\right)\ex^{\im\heffd\left(t_{2}-t_{0}\right)} , \label{eq:term2} \qquad \\
& & \bG_{\left(3\right)}^{<}\left(t_{1},t_{2}\right) \nonumber \\
& = & -\im\ex^{-\im\heff\left(t_{1}-t_{0}\right)}\int_{t_0}^{t_1} \ud\bar{t}\ex^{\im\heff\left(\bar{t}-t_{0}\right)}\left[\bgS^{\rceil}\star\bG^{\lceil}\right]\left(\bar{t},t_{0}\right)\ex^{\im\heffd\left(t_{2}-t_{0}\right)} \nonumber \\
& = & \ex^{-\im\heff\left(t_{1}-t_{0}\right)}\int\frac{\ud\omega}{2\pi}f\left(\omega-\mu\right)\sum_\alpha \bK_{\alpha}\left(t_{1},t_{0};\omega\right)\bgG_{\alpha}\bG^{\text{a}}\left(\omega\right)\ex^{\im\heffd\left(t_{2}-t_{0}\right)} \label{eq:term3} .
\eeq
There is also a term due to the real-time convolution along horizontal
branches of the contour only:
\beq
& & \bG_{\left(4\right)}^{<}\left(t_{1},t_{2}\right) \nonumber \\
& = & \im\ex^{-\im\heff\left(t_{1}-t_{0}\right)}\int_{t_0}^{t_2}\ud\bar{t}\ex^{\im\heff\left(t_{1}-t_{0}\right)}\left[\bG^{\text{r}}\cdot\bgS^{<}\right]\left(t_{1},\bar{t}\right)\ex^{-\im\heffd\left(\bar{t}-t_{0}\right)}\ex^{\im\heffd\left(t_{2}-t_{0}\right)} \nonumber \\
& = & \ex^{-\im\heff\left(t_{1}-t_{0}\right)}\int\frac{\ud\omega}{2\pi}f\left(\omega-\mu\right)\sum_\alpha \im\bK_{\alpha}\left(t_{1},t_{0};\omega\right)\bgG_{\alpha}\bK_{\alpha}^{\dagger}\left(t_{2},t_{0};\omega\right)\ex^{\im\heffd\left(t_{2}-t_{0}\right)} . \nonumber \\ \label{eq:term4}
\eeq
Expanding the compact expression~\eqref{eq:greater/lesserGF}, one easily sees that
it is equal to the sum of the $\bG_{\left(i\right)}^{<}\left(t_{1},t_{2}\right)$.

\begin{figure}[t]
\centering
\includegraphics[width=0.35\textwidth]{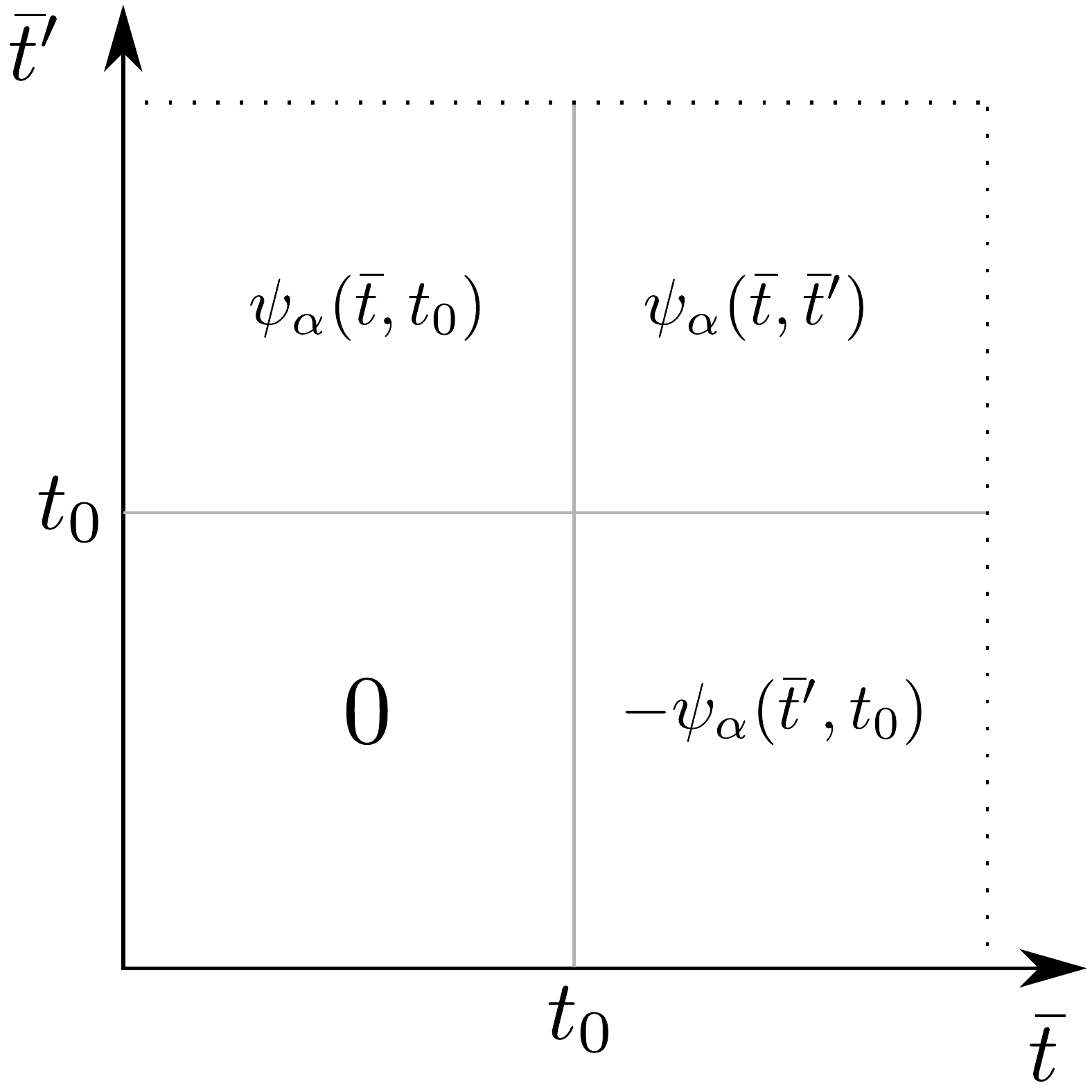}
\caption{Visualization of the double-time integration in Eq.~\eqref{eq:jauhodyson}.}
\label{fig:twotimeplane}
\end{figure}

An alternative approach to the derivation of this formalism is to
work on the original Keldysh contour running on an upper branch from
$-\infty$ to the measurement time $t>t_{0}$ and a lower branch from
$t$ back to $-\infty$~\cite{Keldysh1965}. We then use the Dyson equation
for the lesser GF, as in Ref. \cite{Jauho1994}:
\begin{equation}
\bG^{<}\left(t_{1},t_{2}\right)=\int_{-\infty}^{t} \ud\bar{t}\int_{-\infty}^{t} \ud\bar{t}'\bG^{\text{r}}\left(t_{1},\bar{t}\right)\bgS^{<}\left(\bar{t},\bar{t}'\right)\bG^{\text{a}}\left(\bar{t}',t_{2}\right) . \label{eq:jauhodyson}
\end{equation}
To account for the presence of the lead--molecule coupling in the partition-free
approach, we utilize formulas for the retarded and advanced Green's
functions that are valid before and after $t_{0}$:
\beq
\bG^{\text{r}}\left(t_{1},t_{2}\right) & = & -\im\theta\left(t_{1}-t_{2}\right)\ex^{-\im\heff\left(t_{1}-t_{2}\right)} , \\
\bG^{\text{a}}\left(t_{1},t_{2}\right) & = & \im\theta\left(t_{2}-t_{1}\right)\ex^{-\im\heffd\left(t_{1}-t_{2}\right)} .
\eeq
This is equivalent to using the Keldysh contour hamiltonian specified in Eqs. (2)-(4) with the coupling time pushed to the distant past ($t_{c} \rightarrow
  -\infty$) whilst the bias switch-on time $t_{0}$ remains finite ($t_{c} \neq t_{0}$). Note that information on the coupling is included in these functions
via the effective hamiltonian $\heff$. All of the information
about the bias switch on process is then contained in the lesser self-energy,
which has the form \cite{Ridley2015}:
\begin{equation}
\bgS_{\alpha}^{<}\left(\bar{t},\bar{t}'\right)=\im\bgG_{\alpha}\ex^{-\im\psi_{\alpha}\left(\bar{t},\bar{t}'\right)}\int\frac{\ud\omega}{2\pi}f\left(\omega-\mu\right)\ex^{-\im\omega\left(\bar{t}-\bar{t}'\right)} . \label{eq:sigless}
\end{equation}
Recall that the bias is specified as follows:
\be
V_{\alpha}\left(t\right)= \begin{cases}
V_{\alpha}\left(t\right), & t\geq t_{0}\\
0, & t<t_0 \end{cases}
\ee
where the functional form of $V_{\alpha}\left(t\right)$ is completely
arbitrary. Then we can split the integral in Eq. (\ref{eq:jauhodyson})
into integrals over four different quadrants of the two-time plane,
each with a different phase factor $\psi_{\alpha}\left(\bar{t},\bar{t}'\right)$, see Fig.~\ref{fig:twotimeplane}. These phase factors then enter Eq.~\eqref{eq:sigless}. We write the four quadrants for Eq.~\eqref{eq:jauhodyson} as
\be
\bG^{<}\left(t_{1},t_{2}\right)=\sum_{i=1}^4 I_{i} .
\ee

\textbf{Quadrant 1:} $\bar{t}<t_{0},\,\bar{t}'<t_{0}$, $\psi_{\alpha}\left(\bar{t},\bar{t}'\right)=0$
\beq
I_{1} & = & \int_{-\infty}^{t_0} d\bar{t}\int_{-\infty}^{t_0} d\bar{t}'\bG^{\text{r}}\left(t_{1},\bar{t}\right)\bgS^{<}\left(\bar{t},\bar{t}'\right)\bG^{\text{a}}\left(\bar{t}',t_{2}\right)\label{eq:jauhodyson-1} \\
& = & \int_{-\infty}^{t_0} d\bar{t}\int_{-\infty}^{t_0} d\bar{t}'\ex^{-\im\heff\left(t_{1}-\bar{t}\right)}\int\frac{\ud\omega}{2\pi}f\left(\omega-\mu\right)\sum_\alpha \im\bgG_{\alpha}\ex^{-\im\omega\left(\bar{t}-\bar{t}'\right)} \ex^{\im\heffd\left(\bar{t}'-t_{2}\right)}\nonumber \\
& = & \ex^{-\im\heff\left(t_{1}-t_{0}\right)}\int\frac{\ud\omega}{2\pi}f\left(\omega-\mu\right)\sum_\alpha \im\bA_{\alpha}\left(\omega\right)\ex^{\im\heffd\left(t_{2}-t_{0}\right)} \nonumber \\
& = & \bG_{\left(1\right)}^{<}\left(t_{1},t_{2}\right) .
\eeq
On the last line we identified the same result from Eq.~\eqref{eq:term1}. One can easily obtain similar results for the other quadrants and identify them with Eqs.~\eqref{eq:term2},~\eqref{eq:term3} and~\eqref{eq:term4}.

\textbf{Quadrant 2:} $\bar{t}<t_{0},\,\bar{t}'\geq t_{0}$, $\psi_{\alpha}\left(\bar{t},\bar{t}'\right)=\psi_{\alpha}\left(\bar{t},t_{0}\right)$
\begin{equation}
I_{2}=\int_{-\infty}^{t_0} d\bar{t}\int_{t_0}^{t_2}\ud\bar{t}'\bG^{\text{r}}\left(t_{1},\bar{t}\right)\bgS^{<}\left(\bar{t},\bar{t}'\right)\bG^{\text{a}}\left(\bar{t}',t_{2}\right)=\bG_{\left(2\right)}^{<}\left(t_{1},t_{2}\right) . \label{eq:jauhodyson-1-1}
\end{equation}

\textbf{Quadrant 3:} $\bar{t}\geq t_{0},\,\bar{t}'<t_{0}$, $\psi_{\alpha}\left(\bar{t},\bar{t}'\right)=-\psi_{\alpha}\left(\bar{t}',t_{0}\right)$
\begin{equation}
I_{3}=\int_{t_0}^{t_1} \ud\bar{t}\int_{-\infty}^{t_0} d\bar{t}'\bG^{\text{r}}\left(t_{1},\bar{t}\right)\bgS^{<}\left(\bar{t},\bar{t}'\right)\bG^{\text{a}}\left(\bar{t}',t_{2}\right)=\bG_{\left(3\right)}^{<}\left(t_{1},t_{2}\right) . \label{eq:jauhodyson-1-1-1}
\end{equation}

\textbf{Quadrant 4:} $\bar{t},\,\bar{t}'\geq t_{0}$, $\psi_{\alpha}\left(\bar{t},\bar{t}'\right)=\psi_{\alpha}\left(\bar{t},\bar{t}'\right)$
\begin{equation}
I_{4}=\int_{t_0}^{t_1} \ud\bar{t}\int_{t_0}^{t_2}\ud\bar{t}'\bG^{\text{r}}\left(t_{1},\bar{t}\right)\bgS^{<}\left(\bar{t},\bar{t}'\right)\bG^{\text{a}}\left(\bar{t}',t_{2}\right)=\bG_{\left(4\right)}^{<}\left(t_{1},t_{2}\right) . \label{eq:jauhodyson-1-1-1-1}
\end{equation}

Thus, there is an exact correspondence between the partition-free
approach of Ref.~\cite{Ridley2015} and the partitioned approach of Ref.~\cite{Jauho1994}. A similar procedure can be used to derive the same current in either approach. In Ref.~\cite{Ridley2015}, the current was obtained in terms of the following convolution integrals:
\be\label{eq:CURRENT}
I_{\alpha}\left(t\right)=-4\textrm{Re}\textrm{Tr}_{C}\left[\bgS_{\alpha}^{<}\cdot\bG^{\text{a}}+\bgS_{\alpha}^{\text{r}}\cdot\bG^{<}+\bgS_{\alpha}^{\rceil}\star\bG^{\lceil}\right]\left(t,t\right) .
\ee
However, one can restrict the integration domain to the Keldysh contour
used in Ref. \cite{Jauho1994} and obtain the same formula as Eq.~(\ref{eq:CURRENT}) from the following:
\be
I_{\alpha}\left(t\right) = -4\textrm{Re}\int_{-\infty}^{t} \ud\bar{t}\textrm{Tr}_{C}\left[\bgS_{\alpha}^{<}\left(t,\bar{t}\right)\bG^{\text{a}}\left(\bar{t},t\right)+\bgS_{\alpha}^{\text{r}}\left(t,\bar{t}\right)\bG^{<}\left(\bar{t},t\right)\right] 
\ee
where we split the lesser self energy into two parts for the $\bar{t}\lessgtr t_{0}$
time regimes and recall that the retarded self-energy in the WBLA
is singular in time, see Eq.~\eqref{eq:wbla}.

\section{Conclusions}

In this contribution, we have shown how to choose a Hamiltonian within the partitioned switch-on approach that exactly replicates a partition-free voltage quench. This shows that the result of Odashima \emph{et al.} is more general than was indicated in their paper, and also provides a first step towards a transient version of the Theorem of Equivalence~\cite{Stefanucci2004}. In future work, we will investigate the validity of this formal mapping beyond the WBLA. In addition, it is currently an open question as to whether the reverse mapping from a partitioned to a partition-free representation is always possible. We will address this topic in a forthcoming paper.

%
%

\begin{acknowledgements}
This work was financially supported by the Raymond and Beverly Sackler Center for Computational Molecular and Materials Science, Tel Aviv University (M.R.) and by the DFG (Grant No. SE 2558/2-1) through the Emmy Noether program (R.T.).
\end{acknowledgements}

\bibliographystyle{spphys}       



\end{document}